\documentstyle[prb,aps,epsf]{revtex}

\begin{document}

\draft

\title{Generalized quasiperiodic Rauzy tilings}
\author{Julien Vidal and R\'{e}my Mosseri}
\address{ Groupe de Physique des Solides, CNRS UMR 7588,\\
Universit\'{e}s Pierre et Marie Curie Paris 6 et Denis Diderot Paris 7,\\
2 place Jussieu, 75251 Paris Cedex 05 France}

\maketitle

\begin{abstract}
We present a geometrical description of new canonical $d$-dimensional 
codimension one
quasiperiodic tilings based on generalized Fibonacci sequences. These 
tilings are made up
of rhombi in $2d$ and rhombohedra in $3d$ as the usual Penrose and 
icosahedral tilings. Thanks to a natural
indexing of the sites according to their local environment, we easily 
write down, for any approximant, the sites
coordinates, the connectivity matrix and we compute the structure factor.
\end{abstract}

\section*{Introduction}

Quasiperiodic tilings have been widely studied over the past decades, their
main interest lying in their guessed relation with real quasicrystalline
alloy\cite{Shechtman} atomic structure. Indeed, most of the specific features
and physical properties associated with quasiperiodic order 
(electronic structure, localized
phason degree of freedom, nature and dynamics of dislocations, ...) 
are already present in
these simplified models. The two standard ways to
generate these tilings are the cut and project method\cite{KD,KKL} which relate
them to selected pieces of higher dimensional lattices, and the grid
algorithm\cite{deBruijn} which insists on a dual picture of 
intersecting lines or
planes.
In both cases, one obtains structures made up of different types of 
tile which are segments in
$1d$, rhombi in $2d$ and rhombohedra in $3d$. The simplest (and most 
famous) examples
correspond to tilings with two differents tiles (the one-dimensional 
Fibonacci chain, the Penrose
(pentagonal)\cite{Penrose} and Amman-Beenker\cite{Beenker} (octogonal)
two-dimensional tilings and the icosahedral $3d$ tiling) although 
other quasiperiodic structures
with more different types of tile have also been proposed. In $d\geq 
2$, all these tilings
display a complex local order with various coordination number.

Concerning the study of the physical properties (spectrum, 
eigenstates, conductivity,...), it has
been proved more efficient to consider simpler systems such as direct 
products of
periodic chains\cite{Zhong,Ashraff,Ueda,Sire_Aussois} embedded in a 
quasiperiodic
potential. The main advantage of these structures is that their characteristics
(sub-ballistic transport, self-similar eigenstates) can be easily 
deduced from the $1d$ case~; on the other
hand, since all the sites have the same coordination number, their 
topology is trivial.
In $2d$, Penrose-like tilings have nevertheless been studied 
revealing the same type of
features\cite{Kohmoto_Sutherland1,Tsunetsugu_conductance,GrimmICQ6,Sch 
reiber,Passaro_Octo,Sire_Bel,Zhong_octo},
but in $3d$, the incidence of the topological quasiperiodicity on the 
electronic properties has been poorly
investigated\cite{Schwabe_Mag_field,Kasner3D} essentially because of 
the geometrical complexity.   In this
context, it would be appreciable to have simple structures 
susceptible to provide a better framework for such
studies.

The aim of this paper is to present new quasiperiodic tilings (of
arbitrary dimension) following from a natural extension of the 
Fibonacci sequence, which allow
for a coherent indexing of the sites while keeping some of the 
interesting self-similar
properties of the Penrose-like tilings. We give the procedure to 
build any approximant of
these topologically non trivial tilings using the conumbering
scheme\cite{Mosseri_conumbering,Mosseri_conumbering2}.
In the quasiperiodic limit, these tilings are closely related to 
those initially
considered by Rauzy\cite{Rauzy} in two dimensions and will therefore be called
generalized Rauzy tilings.   In the first part, we recall the 
principle of the conumbering
scheme that allows to generate all the sites of any codimension one
structures by iterating the so-called generating vector ${\bf g}$. As an
illustration, we apply this scheme to build the well-known Fibonacci 
chain for which it is
possible to exactly determine the coordinates of ${\bf g}$.
In the second part, we generalize these results in higher dimension
and we describe, in details, the characteristics of  the generalized 
Rauzy tilings in $2d$ and
$3d$. The site coordinates are explicitely  obtained in terms of the 
generalized Fibonacci
numbers.  In the third part,  we give the connectivity matrix of 
these canonical tilings that
is relevant for tight-binding electrons problems. The derivation of 
the structure factor is
given in the fourth part.
Finally, we propose, in the appendix, a dual point of view in which the $1d$
quasiperiodic system presented  have a high codimension although
based on the same type of sequence.

%
%
\section{The conumbering scheme~: application to the Fibonacci chain}
%
%
 
In the cut and project algorithm commonly used to build quasiperiodic 
structures, one considers
a $D$-dimensional hypercubic lattice and $d$-dimensional subspace 
that define the ``physical"
space. The $(D-d)$-dimensional subspace define the perpendicular 
space whose dimension is
called the codimension. Thus, $d$-dimensional codimension one tilings 
are generated
from a $(d+1)$-dimensional hypercubic lattice.
The unidimensional character of the perpendicular space of these 
tilings provides a natural
ordering of the sites which amounts to classify them according to 
their local environment~: two
sites with close coordinates in the perpendicular space have a 
similar neighbourhood.
As shown below, each site of a codimension one tiling can actually be 
indexed by a unique coordonnate,
its conumber, that is related to its position in the perpendicular 
space. For any approximant structure,
this is achieved by using a generating vector which, upon simple 
iteration, fully determines the
coordinates of the sites inside the unit 
cell\cite{Mosseri_conumbering,Mosseri_conumbering2}.
Since this method is valid for any approximant, the quasiperiodic 
structure can therefore be approached
asymptotically.

\begin{figure}
\centerline{\epsfxsize=150mm
\epsffile{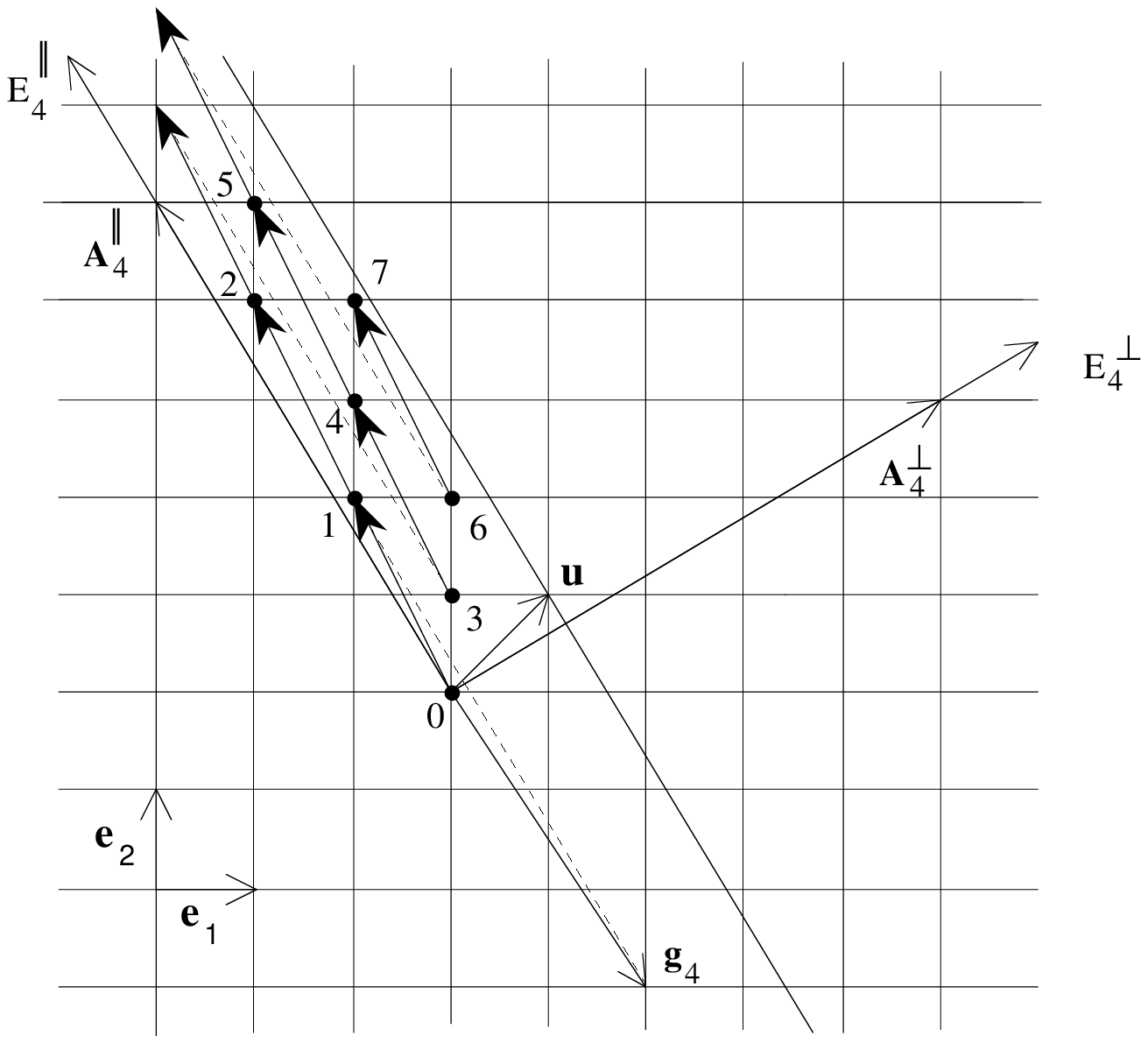}}
\vspace{-115mm}
\caption{Conumbering of the sites of the $4^{th}$ order approximant 
of the Fibonacci chain
  upon iteration of a generating vector.  ${\bf A}_{4}^{\perp}=(5,3)$,
${\bf A}_{4}^{\parallel}=(-3,5)$ and  ${\bf g}_{4}=(2,-3)$.}
\label{Fibo_chain}
\end{figure}

To illustrate this mechanism, we shall focus on the $2\rightarrow 1$ case.
We consider a square lattice, and draw a line through the origin $O$,
denoted $E^{\parallel}$ (the parallel or physical space), of rational slope
$\alpha =p/q$, with $p$ and $q$ mutually prime (see figure 
\ref{Fibo_chain}). The sites
of the approximant structure are obtained by an orthogonal projection on
$E^{\parallel}$ of the square lattice sites contained inside the 
semi-open band generated
by sliding the unit square along $E^{\parallel}$. Since 
$E^{\parallel}$ has rational
slope, it contains a set of regularly spaced sites of the square lattice
which defines a unit cell with $n=|p|+|q|$ sites and a cell vector 
${\bf A}^{\parallel}$.
We define the so-called generating vector ${\bf g}$ as the vector of 
smallest norm\cite{Sign},
joining the origin to the square lattice site closest to 
$E^{\parallel}$ (but not belonging to
$E^{\parallel}$).  It can then be 
shown\cite{Mosseri_conumbering,Mosseri_conumbering2} that the
sites inside the band are obtained by successive translations of 
${\bf g}$ ({\it modulo} ${\bf
A}^{\parallel}$)~:
%
%
\begin{equation}
{\bf r}^j=j\:{\bf g} \:\: modulo \:\: {\bf A^\parallel}\, , \: j\in [0,n-1]
\mbox{ . }
\label{conum_coord}
\end{equation}
%
%
This indexation by a unique coordinate (the conumber) provides a 
natural ordering with respect to
the distance (before projection) from the parallel space, {\it i.~e.} 
with respect to the
local environment. Note that the origin has, by definition, the conumber $0$.
For arbitrary sequences of approximants, the generator coordinates 
are given  by
a specific element of the  Farey tree
decomposition\cite{Mosseri_conumbering,Mosseri_conumbering2} but, for
particular rational slopes, it is possible to explicitely write down 
the generating vector as we
shall now see for the Fibonacci chain.\\

We denote by ${\cal B}_{2}$=$({\bf e}_{1},{\bf e}_{2})$ the canonical 
orthonormal basis of the
square lattice, and consider the vector ${\bf A}_{k}^{\perp}$ whose 
coordinates in ${\cal B}_{2}$
are given by $(F_{k},F_{k-1})$. The index $k$ refers to the 
approximant order and the Fibonacci
sequence $(F_l)_{l\in {\bf Z}}$ is defined as follows~:
%
%
\begin{equation}
F_{l+1}=F_{l}+F_{l-1} \:\:\: \hbox{with} \:\:\: F_0=F_1=1
\mbox{ . }
\label{recur}
\end{equation}
%
%
Usually, this sequence is only defined for $l\geq 0$ but, here, we 
shall also consider
negative values of $l$. As readily seen on the recursion relation (\ref{recur})
${\displaystyle \lim_{n\rightarrow \infty}} F_{n+1}/F_{n}=\tau$ where 
the golden mean
$\tau=(1+\sqrt{5})/2$ is the Pisot\cite{Pisotdef} solution  of the quadratic
equation $x^{2}=x+1$. ${\bf A}_{k}^{\parallel}$ is then defined as 
the vector of smallest
norm\cite{Sign} with integer coordinates perpendicular to ${\bf 
A}_{k}^{\perp}$.
$({\bf A}_{k}^{\parallel},{\bf A}_{k}^{\perp})$ forms a basis of the 
so-called trace lattice
$\Lambda _{k}$. One also defines the band lattice $\Sigma_{k}$ 
generated by ${\bf
A}_{k}^{\parallel}$ and the vector ${\bf u}=(1,1)$ (see figure 
(\ref{Fibo_chain}) for $k=4$).
Let $L_{k}$ (resp. $S_{k}$) be the matrix transforming the canonical basis
${\cal B}_{2}$ into the $\Lambda _{k}$ (resp. $\Sigma _{k}$) basis. 
The number of sites
of the  square lattice contained in the $ \Lambda _{k}$ (resp. $S_{k}$)
unit cell thus reads $l_{k}=|\det L_{k}|$ (resp. $s_{k}=|\det S_{k}|$).
To determine the coordinate of the generating vector ${\bf g}_{k}$, 
we introduce
the matrix~$M$~:
%
%
\begin{eqnarray}
M=\left( \begin{array}{cc}
                       1 & 1 \\
                       1 & 0
           \end{array}
          \right)
\mbox{ , }
\end{eqnarray}
%
%
that verifies the golden mean equation~:
%
%
\begin{equation}
M^{2}=M+1
\mbox{ . }
\end{equation}
%
%
%
%
%
%

\noindent The powers of $M$ can be expressed in terms of the
Fibonacci numbers~:
%
%
\begin{eqnarray}
M^{n}=\left(
					      \begin{array}{ll}
                   F_{n} & F_{n-1} \\
                   F_{n-1} & F_{n-2}
            \end{array}
       \right) ,
\qquad \forall n\in {\bf Z}.
\label{mat.Fibo}
\end{eqnarray}
%
%
Note that we also allow the index $n$ to be negative.
Since the first column of $M^{k}$ represents the coordinates of ${\bf 
A}_{k}^{\perp}$, it comes
immediately that the second line of $M^{-k}$ gives the coordinates of
${\bf A}_{k}^{\parallel}$ in the basis ${\cal B}_{2}$~:
%
%
\begin{equation}
{\bf A}_{k}^{\parallel}=(F_{-k-1},F_{-k-2})\mbox{ . }
\end{equation}
%
%
The matrices $L_{k}$ and $S_{k}$ thus reads~:
%
%
\begin{equation}
L_{k}=\left(
\begin{array}{ll}
F_{-k-1} & F_{k} \\
F_{-k-2} & F_{k-1}
\end{array}
\right) \:\:\:\mbox{ and } \:\:\:S_{k}=\left(
\begin{array}{lr}
F_{-k-1} & 1 \\
F_{-k-2} & 1
\end{array}
\right)
\mbox{ . }
\end{equation}
%
%

To determine the number of sites in the $\Sigma_{k}$ unit cell,
one can either directly compute $s_{k}$ or one can remark that this
number is also given by the scalar product of ${\bf u}$ and ${\bf 
A}_{k}^{\perp}$.
One thus readily obtains~:
%
%
\begin{equation}
s_{k}={\bf u}.{\bf A}_{k}^{\perp}=F_{k}+F_{k-1}=F_{k+1}\mbox{ . }
\end{equation}
%
%

To determine ${\bf g}_{k}$, one uses the fact that the quadrilateral
generated by ${\bf A}_{k}^{\parallel}$ and ${\bf g}_{k}$ should have a unit
area since it cannot contain a lattice site (except on its 
boundary). Since ${\bf
A}_{k}^{\parallel}$ corresponds to the second line of $M^{-k}$ and 
since $\det M=-1$, the first
line of $M^{-k}$ can be chosen to be the coordinates (in the basis 
${\cal B}_{2})$ of the
generating vector~:
%
%
\begin{equation}
{\bf g}_{k}=(F_{-k},F_{-k-1})
\end{equation}
%
%

The coordinates of the sites contained in the $\Sigma_{k}$ unit cell
(sites of the approximant structure before projection), read in the 
basis ${\cal
B}_{2}$~:
%
%
\begin{equation}
{\bf r}^j_k=j\:{\bf g}_k \:\: modulo \:\: {\bf A}_{k}^{\parallel}\,, 
\: j\in [0,s_k-1]
\mbox{ . }
\end{equation}
%
%
The $modulo$ operator allows to carry back all the sites in the same
elementary cell of the band lattice.
In the basis $({\bf A}_{k}^{\parallel},{\bf A}_{k}^{\perp})$ \ of the 
trace lattice
$\Lambda _{k}$, the coordinates finally write~:
%
%
\begin{equation}
{\bf r}_{k}^{j}=\mbox{Frac}\,(j\,L_{k}^{-1}\,{\bf g}_{k})\,,\;j\in \lbrack
0,s_{k}-1]\mbox{ , }  \label{modulo1}
\end{equation}
%
%
where $\mbox{Frac}\,({\bf r})$ represents the fractional part of the ${\bf r}$
coordinates. The interest of introducing the trace lattice is that the
coordinates of the sites after projection onto $E_k^{\parallel}$ are
readily given by the first component of ${\bf r}_{k}^{j}$
expressed in the $({\bf A}_{k}^{\parallel},{\bf A}_{k}^{\perp})$ basis.

In the next section, we propose to generalize the above
construction  to codimension one structures in any dimension, such that the
perpendicular direction is related to generalized Fibonacci numbers . These
numbers are obtained from the Pisot solution  of the polynomial equation
$x^{D}=\sum_{j=0}^{D-1}x^{j}$, the $D=2$ case giving the golden mean.
%
%
\section{ The generalized Rauzy tilings}
%
%
 
The first geometrical construction based on the roots of this 
equation (for $D=3$) was
proposed in 1982 by the mathematician G. Rauzy\cite{Rauzy}.
Detailed analyses of these original tilings focusing on their self-similar
properties and their fractal boundaries can be found in 
Refs.\cite{Hito_Rauzy,Messaoudi,Arnoux}.
Nevertheless, the tilings that we describe thereafter are different 
although close to those
initially studied. Indeed, the construction proposed by Rauzy does 
not rely on the standard
cut and project algorithm since the sites of the cubic lattice chosen 
for the projection step
are contained in a cylinder centered around the perpendicular 
direction and not in the band
lattice. In addition, the  construction of the approximant structures 
is completely new.\\

In order to show the simplicity of the generalization, we will use, 
in the following, the same notations for the
$D\rightarrow D-1$ tilings as for the $2\rightarrow 1$ tilings.

%
%
\subsection{The two-dimensional case}
%
%

As for the $2\rightarrow 1$ case, we shall use the conumbering scheme 
to generate
the $3\rightarrow 2$ approximant  structures.
We endow the standard cubic lattice of  a canonical orthonormal basis
${\cal B}_{3}=({\bf e}_{1},{\bf e}_{2},{\bf e}_{3})$ and, by
analogy with the Fibonacci chain, we choose as perpendicular space 
the direction
defined by vector ${\bf A}_{k}^{\perp}=(F_{k},F_{k-1},F_{k-2})$ where 
the generalized
Fibonacci sequence $(F_{l})_{l\in {\bf Z}}$ is defined as follows~:
%
%
\begin{equation}
F_{l+1}=F_{l}+F_{l-1}+ F_{l-2} \:\:\: \hbox{with} \:\:\: F_{-1}=0,\, F_0=F_1=1
\mbox{ . }
\label{recurRauzy}
\end{equation}
%
%
As previously, the ratio of two successive elements of this sequence 
converges toward an
irrational limit~:
%
%
\begin{equation}
\lim_{n\rightarrow \infty} F_{n+1}/F_{n}=\alpha
={\frac{4+\left( 19+3\sqrt{33}\right)^{1/3}+\left( 19+3\sqrt{33}
\right) ^{2/3}}{3\left( 19+3\sqrt{33}\right) ^{1/3}}}\simeq 1.83929
\mbox{ , }
\end{equation}
%
%
where $\alpha$ is the Pisot root of the cubic equation $x^{3}=x^{2}+x+1$.
We introduce the matrix $M$~:
%
%
\begin{eqnarray}
M=\left( \begin{array}{ccc}
                       1 & 1 & 1 \\
                       1 & 0 & 0 \\
                       0 & 1 & 0
           \end{array}
          \right)
\mbox{ , }
\end{eqnarray}
%
%
which satisfies $M^{3}=M^{2}+M+1$, and whose successive powers read~:
%
%
\begin{equation}
M^{n}=\left(
\begin{array}{lll}
F_{n} & F_{n-1}+F_{n-2}   & F_{n-1} \\
F_{n-1}   & F_{n-2}+F_{n-3} & F_{n-2} \\
F_{n-2} & F_{n-3}+F_{n-4} & F_{n-3}
\end{array}
\right) \qquad \forall n\in {\bf Z}
\mbox{ . }
\end{equation}
%
%
Below are shown the first (positive and negative) powers of $M$ which 
display a remarkable
pattern on both parts of the identity matrix $ M^0$.
%
%
\begin{equation}
\begin{array}{l}
M^{4} \\
\\
M^{2} \\
\\
M^{0} \\
\\
M^{-2} \\
\\
M^{-4}
\end{array}
\begin{array}{rrr}
7 & 6 & 4 \\
4 & 3 & 2 \\
2 & 2 & 1 \\
1 & 1 & 1 \\
{\bf 1} & {\bf 0} & {\bf 0} \\
{\bf 0} & {\bf 1} & {\bf 0} \\
{\bf 0} & {\bf 0} & {\bf 1} \\
1 & -1 & -1 \\
-1 & 2 & 0 \\
0 & -1 & 2 \\
2 & -2 & -3
\end{array}
\hspace{3mm}
\begin{array}{l}
M^{3} \\
\\
M \\
\\
M^{-1} \\
\\
M^{-3}
\end{array}
\label{tab}
\end{equation}
%
%
Note that the vertical sequences of numbers, in each column, obey relation
(\ref{recurRauzy}) from bottom to top.
To build the $k^{th}$ order approximant, one remarks that ${\bf A}_{k}^{\perp}$
corresponds to the first column of $M^{k}$ so that the two vectors
${{\bf A}_{k}^{\parallel}}^{1}$ and ${{\bf A}_{k}^{\parallel}}^{2}$ 
which generate the
parallel space are directly obtained from the second and third line 
of $M^{-k}$~:
%
%
\begin{eqnarray}
{{\bf A}_{k}^{\parallel}}^1&=&(F_{-k-1},F_{-k-2}+F_{-k-3},F_{-k-2})\\
\nonumber\\
{{\bf A}_{k}^{\parallel}}^2&=&(F_{-k-2},F_{-k-3}+F_{-k-4},F_{-k-3})
\mbox{.}
\end{eqnarray}
%
%
$({{\bf A}_{k}^{\parallel}}^{1},{{\bf A}_{k}^{\parallel}}^{2},{\bf 
A}_{k}^{\perp})$
defines a basis of the trace lattice $\Lambda _{k}$~; the band 
lattice $\Sigma _{k}$ is
generated by $({{\bf A}_{k}^{\parallel}}^{1},{{\bf A}_{k}^{\parallel}}^{2})$
and the vector ${\bf u}=(1,1,1)$, which joins the origin to the 
extremity of the unit cube
whose projection onto ${\bf A}_{k}^{\perp}$ has the highest positive magnitude.
The number of sites in a $\Sigma _{k}$ unit cell is given by~:
%
%
\begin{equation}
s_{k}={\bf u}.{\bf A}_{k}^{\perp}=F_{k}+F_{k-1}+F_{k-2}=F_{k+1}
\mbox{.}
\end{equation}
%
%
The generating vector ${\bf g}_{k}$ is determined by the condition~:
$\det \left({{\bf A}_{k}^{\parallel}}^{1},{{\bf 
A}_{k}^{\parallel}}^{2},{\bf g}_{k}\right)=1$.
Since $\det (M^{-k})=(\det M)^{-k}=1$, one can identify the coordinates of
${\bf g}_{k}$ with the first line of $M^{-k}$~:
%
%
\begin{equation}
{\bf g}_{k}=(F_{-k},F_{-k-1}+F_{-k-2},F_{-k-1})
\mbox{.}
\end{equation}
%
%
and the coordinates (before projection) of the sites contained in a 
$\Sigma _{k}$
unit cell read in the basis ${\cal B}_{3}$~:
%
%
\begin{equation}
{\bf r}^j_k=j\:{\bf g}_k \:\: modulo \:\: ({{\bf A}_{k}^{\parallel}}^1,
{{\bf A}_{k}^{\parallel}}^2)
\, , \: j\in [0,s_k-1]
\mbox{ . }
\label{babar}
\end{equation}
%
%
It is still interesting to express these coordinates in the trace lattice
basis $({{\bf A}_{k}^{\parallel}}^{1},{{\bf 
A}_{k}^{\parallel}}^{2},{\bf A}_{k}^{\perp})$~:

%
%
\begin{equation}
{\bf r}_{k}^{j}=\mbox{Frac}\,(j\,L_{k}^{-1}{\bf g}_{k})\,,\;j\in \lbrack
0,s_{k}-1]\mbox{ , }
\label{expression}
\end{equation}
%
%
where~:
%
%
\begin{equation}
L_{k}=\left(
\begin{array}{lll}
F_{-k-1} & F_{-k-2} & F_{k} \\
F_{-k-2}+F_{-k-3} & F_{-k-3}+F_{-k-4} & F_{k-1} \\
F_{-k-2} & F_{-k-3} & F_{k-2}
\end{array}
\right) \mbox{ . }
\end{equation}
%
%

\begin{figure}
\centerline{\epsfxsize=150mm
\hspace{50mm}
\epsffile{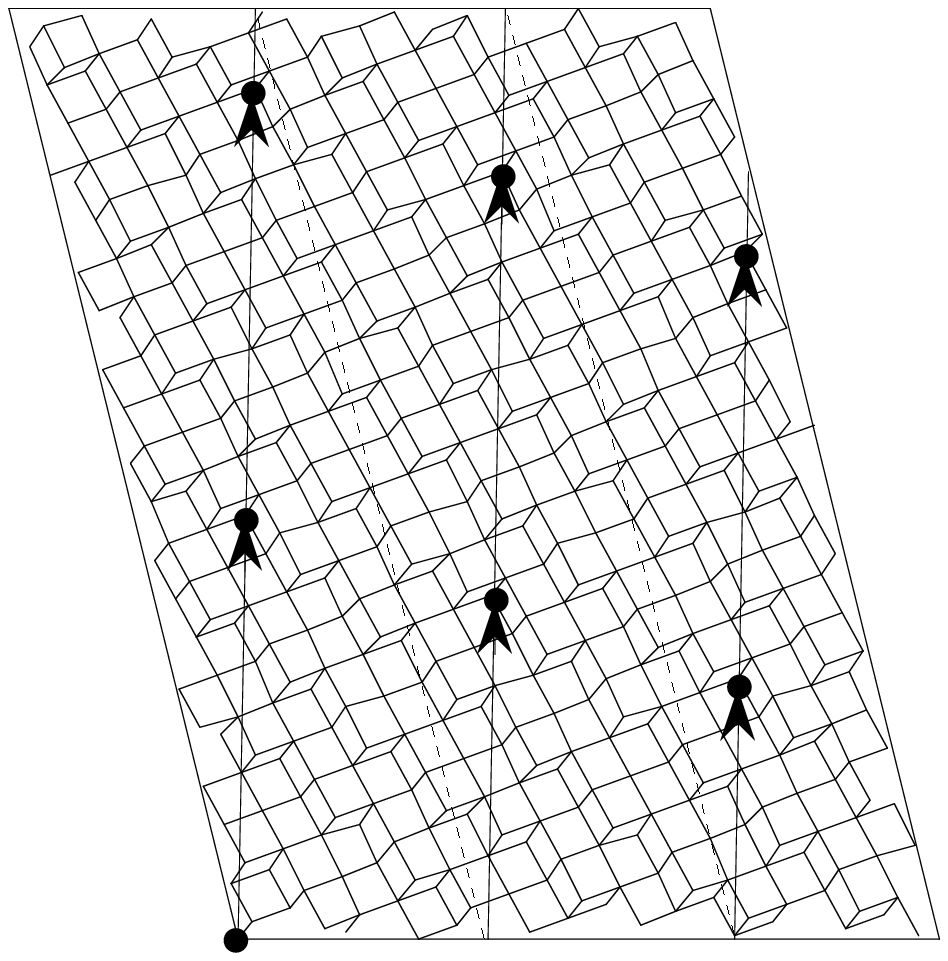}}
\vspace{-130mm}
\caption{Unit cell of the $10^{th}$ order approximant (504 sites). The first
iterations of the generator are shown.}
\label{Rauzy_app}
\end{figure}

After projection onto $E^{\parallel}_k$, the $j^{th}$ site coordinates,
in the basis $({{\bf A}_{k}^{\parallel}}^{1},{{\bf 
A}_{k}^{\parallel}}^{2})$ are given by the two
first components of ${\bf r}_{k}^{j} $.
Figure \ref{Rauzy_app} displays an elementary cell of the $10^{th}$ 
order approximant.
%
%
%
%

In the initial cubic lattice, each site is six-fold coordinated so 
that, after projection, the coordination
number $z$ is such that $3\leq z\leq 6$. However, as discussed below, 
the tilings generated here only have sites
with $3\leq z\leq 5$.

%
%
\subsection{The higher dimensional case}
%
%

\hspace{5mm} The simplicity of the above construction results directly from the
particular properties of the generalized Fibonacci numbers. It can therefore
easily extended to codimension one tilings of any higher dimension. 
Thereafter, we briefly
describe the three-dimensional case and the $d$-dimensional case.

The construction in the $4\rightarrow 3$ case is based on the Pisot 
root of the equation~:
%
%
\begin{equation}
x^{4}=x^{3}+x^{2}+x+1
\mbox{ , }
\label{R3eq}
\end{equation}
%
%
whose approximants define the perpendicular space direction in the $4D$
space.

The associated generalized Fibonacci sequence $(F_{l})_{l\in {\bf 
Z}}$ is defined by~:
%
%
\begin{equation}
F_{l+1}=F_{l}+F_{l-1}+F_{l-2}+F_{l-3} \:\:\: \hbox{with} \:\:\:
F_{-1}=F_{-2}=0,\,F_{0}=F_{1}=1
\mbox{ , }
\end{equation}
%
%
and the $M$ matrix satisfying (\ref{R3eq}) reads~:
%
%
\begin{eqnarray}
M=\left( \begin{array}{cccc}
                       1 & 1 & 1 & 1\\
                       1 & 0 & 0 & 0\\
                       0 & 1 & 0 & 0\\
                       0 & 0 & 1 & 0\\
           \end{array}
          \right)
\mbox{ , }
\end{eqnarray}
%
%
so that~:
%
%
\begin{equation}
M^{n}=\left(
\begin{array}{llll}
F_{n}    & F_{n-1}+F_{n-2}+F_{n-3}  & F_{n-1}+F_{n-2} & F_{n-1} \\
F_{n-1}  & F_{n-2}+F_{n-3}+F_{n-4}  & F_{n-2}+F_{n-3} & F_{n-2} \\
F_{n-2}  & F_{n-3}+F_{n-4}+F_{n-5}  & F_{n-3}+F_{n-4} & F_{n-3} \\
F_{n-3}  & F_{n-4}+F_{n-5}+F_{n-6}  & F_{n-4}+F_{n-5} & F_{n-4} \\
\end{array}
\right) \qquad \forall n\in {\bf Z}
\mbox{ . }
\end{equation}
%
%

The $k^{th}$ order approximant is obtained by choosing the vector
${\bf A}_{k}^{\perp}=(F_{k},F_{k-1},F_{k-2},F_{k-3})$ whose 
coordinates are given by the
first column of $M^{k}$. The three parallel space vectors reads~:
%
%
\begin{eqnarray}
{{\bf 
A}_{k}^{\parallel}}^1&=&(F_{-k-1},F_{-k-2}+F_{-k-3}+F_{-k-4},F_{-k-2}+ 
F_{-k-3}, F_{-k-2})\\
{{\bf 
A}_{k}^{\parallel}}^2&=&(F_{-k-2},F_{-k-3}+F_{-k-4}+F_{-k-5},F_{-k-3}+ 
F_{-k-4}, F_{-k-3})\\
{{\bf A}_{k}^{\parallel}}^3&=&(F_{-k-3},F_{-k-4}+F_{-k-5}+F_{-k-6},F_{-k-4}+F_{-k-5}, F_{-k-4} 
)
\mbox{ , }
\end{eqnarray}
%
%
that are respectively the second, third and fourth line of $M^{-k}$.
The band lattice $\Sigma _{k}$ is generated by
$({{\bf A}_{k}^{\parallel}}^{1},{{\bf A}_{k}^{\parallel}}^{2},{{\bf 
A}_{k}^{\parallel}}^{3})$
and the vector ${\bf u}=(1,1,1,1)$. The number of sites in the 
$\Sigma _{k}$ unit cell (and therefore in the
approximant unit cell) is given by~:
%
%
\begin{equation}
s_{k}={\bf u}.{\bf A}_{k}^{\perp}=F_{k}+F_{k-1}+F_{k-2}+F_{k-3}=F_{k+1},
\end{equation}
%
%
%
and the coordinates of the generating vector  are given by the first 
line of $M^{-k}$~:
%
%
\begin{equation}
{\bf g}_k=(F_{-k}, F_{-k-1}+F_{-k-2}+F_{-k-3}, F_{-k-1}+F_{-k-2}, F_{-k-1})
\mbox{ , }
\end{equation}
%
%
The sites coordinates have an expression similar to 
Eq.(\ref{expression}) where $L_k$ is, as before, built from
the coordinates of
${{\bf A}_{k}^{\parallel}}^{1},{{\bf A}_{k}^{\parallel}}^{2},{{\bf 
A}_{k}^{\parallel}}^{3}$ and
${\bf A}_{k}^{\perp}$.\\

The generalization in any dimension $D$ is straightforward considering the
Pisot root of the equation~:
%
%
\begin{equation}
x^{D}=\sum_{j=0}^{D-1}x^{j}\mbox{ , }
\label{RDeq}
\end{equation}
%
%
The remaining of the construction follows mechanically.\\

Let us end this section by remarking that nothing prevents, in any
dimension, to map the whole structure along a direction such that all tiles
(rhombus in $2d$, rhombohedra in $3d$) are identical up to
rotations. Indeed, once the sites have been generated with a given 
choice of the
perpendicular space, this can be achieved by projecting perpendicularly to the
direction of the vector ${\bf u}$ whose each coordinate equals 1. In 
that case, all the
sites are mapped onto the full set of simple lattice vertices 
(triangular in $2d$ and centered
cubic in $3d$) with missing edges. We are then left with a 
quasiperiodic decoration of a periodic
structure, which has the same connectivity but a different geometry 
as the above discussed
tilings. Such deformations might be useful in some context, like for 
instance when one aims
to build row-by-row or plane-by-plane transfer matrices, in which case the
preexisting reticular stratification of the lattice can be very helpful.

%
%
\section{Connectivity matrix of codimension one tilings}
%
%
In this section we write down the so-called connectivity or adjacency 
matrix $K$
of codimension one tilings. This matrix is defined by the following 
rule~: $K_{ij}=1$ if
sites $i$ and $j$ are nearest neighbours ({\it i.~e.} connected by an 
edge) and zero
otherwise.
Such a matrix is of special interest for the study of the electronic 
or phononic
excitations in tight-binding like approaches. In this case, the 
hamiltonian is, indeed,  simply
proportional to K. Being able to easily obtain  such a matrix  may 
therefore  be highly
valuable.

The idea consists in labelling the sites with respect to their 
conumber. As discussed in section
II,  this indexation  classifies the sites according to their 
coordinate in the perpendicular
space direction, or equivalently, in terms of their local environment.
This implies that the conumber difference between two nearest 
neighbour sites along a given
direction is a constant $modulo$ the number of sites in the unit 
cell.  This difference is
readily obtained from the projection onto the perpendicular space of 
each  edges meeting at any
site of the initial hypercubic lattice. It is therefore sufficient to 
determine, for a given site,
the conumber of its nearest neighbours, to write down the full 
connectivity matrix as a band
(T{\oe}plitz-like) matrix.

We shall first illustrate this method with a Fibonacci chain 
approximant, defined
by the perpendicular space vector ${\bf A}_{k}^{\perp}=(F_{k},F_{k-1})$ and
determine the conumber of the nearest neighbours of the origin. In the
canonical basis ${\cal B}_2$ of the square lattice, these sites have 
the following
coordinates $(0,\pm 1)$ and $(\pm 1,0)$. Their projection onto the 
perpendicular space is
given (up to a sign), in the trace lattice $\Lambda _{k}$ basis, by 
${\bf A}_{k}^{\perp}/l_{k}$.
In addition, and by construction, one has ${\bf A}_{k}^{\perp}.{\bf 
g}_{k}=1/l_{k}$ in the
$\Lambda _{k}$ basis. One therefore deduces that the conumber of the 
sites which are nearest
neighbours of the origin (once carried back in a unique unit cell) 
reads $F_{k}$ et
$F_{k-1}$.

As an example, we give the connectivity matrix $K$ of the $4^{th}$ 
order approximant of the Fibonacci
chain displayed in figure (\ref{Fibo_chain}) with periodic boundary conditions:
%
%
\begin{equation}
K=\left(
\begin{array}{cccccccc}
0 & 0 & 0 & 1 & 0 & 1 & 0 & 0 \\
0 & 0 & 0 & 0 & 1 & 0 & 1 & 0 \\
0 & 0 & 0 & 0 & 0 & 1 & 0 & 1 \\
1 & 0 & 0 & 0 & 0 & 0 & 1 & 0 \\
0 & 1 & 0 & 0 & 0 & 0 & 0 & 1 \\
1 & 0 & 1 & 0 & 0 & 0 & 0 & 0 \\
0 & 1 & 0 & 1 & 0 & 0 & 0 & 0 \\
0 & 0 & 1 & 0 & 1 & 0 & 0 & 0
\end{array}
\right) \mbox{ . }
\end{equation}
%
%
There are indeed $F_{5}=8$ sites in the elementary cell and the
conumber of the origin neighbours are $F_{4}=5$ and $F_{3}=3$ (see 
fig. \ref{Fibo_chain}).

The above discussion applies to the generalized Rauzy tilings~: their 
connectivity matrix remains a band matrix
once the sites have been ordered according to their conumber. The 
conumbers of the origin's nearest neighbours
are still given by the coordinates of ${\bf A}_{k}^{\perp}$ which are 
expressed in terms of the generalized
Fibonacci numbers.
We display below the connectivity matrix of the $4^{th}$ order 
approximant of the
two-dimensional generalized Rauzy tiling, with periodic boundary conditions~:
%
%
\begin{equation}
K=\left(
\begin{array}{ccccccccccccc}
0 & 0 & 1 & 0 & 1 & 0 & 0 & 1 & 0 & 0 & 0 & 0 & 0 \\
0 & 0 & 0 & 1 & 0 & 1 & 0 & 0 & 1 & 0 & 0 & 0 & 0 \\
1 & 0 & 0 & 0 & 1 & 0 & 1 & 0 & 0 & 1 & 0 & 0 & 0 \\
0 & 1 & 0 & 0 & 0 & 1 & 0 & 1 & 0 & 0 & 1 & 0 & 0 \\
1 & 0 & 1 & 0 & 0 & 0 & 1 & 0 & 1 & 0 & 0 & 1 & 0 \\
0 & 1 & 0 & 1 & 0 & 0 & 0 & 1 & 0 & 1 & 0 & 0 & 1 \\
0 & 0 & 1 & 0 & 1 & 0 & 0 & 0 & 1 & 0 & 1 & 0 & 0 \\
1 & 0 & 0 & 1 & 0 & 1 & 0 & 0 & 0 & 1 & 0 & 1 & 0 \\
0 & 1 & 0 & 0 & 1 & 0 & 1 & 0 & 0 & 0 & 1 & 0 & 1 \\
0 & 0 & 1 & 0 & 0 & 1 & 0 & 1 & 0 & 0 & 0 & 1 & 0 \\
0 & 0 & 0 & 1 & 0 & 0 & 1 & 0 & 1 & 0 & 0 & 0 & 1 \\
0 & 0 & 0 & 0 & 1 & 0 & 0 & 1 & 0 & 1 & 0 & 0 & 0 \\
0 & 0 & 0 & 0 & 0 & 1 & 0 & 0 & 1 & 0 & 1 & 0 & 0
\end{array}
\right) \mbox{ . }
\label{connect2}
\end{equation}
%
%
One has $F_{5}=13$ sites in the unit cell, and the conumber of the 
origin's nearest neighbours
are $F_{4}=7$, $F_{3}=4$, and $F_{2}=2$.
One can therefore appreciate the advantage of this family of tilings that
provides topologically non trivial structures  which are nevertheless 
very easy to construct
and to encode. Especially in $3d$, this is undoubtlessly one of the 
simplest type of
connectivity matrix that one could expect for quasiperiodic structure 
approximant of any order.\\

The last point we would like to discuss concerns the proportion of 
the different types of site
and their coordination number. Clearly, the coordination number of site
$j$ is the number of $1$ in the $j^{th}$ line  of the connectivity 
matrix $K$. Hence, in the
above example one finds four $3$-fold coordinated sites , five 
$4$-fold coordinated sites, and
four $5$-fold coordinated sites but no $6$-fold coordinated sites.
It is easy to show that the condition to have sites of maximal 
coordination number
$z=2D$ in a generalized Rauzy tiling of type $D\rightarrow (D-1)$ reads~:
%
%
\begin{equation}
F_{k+1}-2F_{k}>0\Leftrightarrow {\frac{F_{k+1}}{F_{k}}}>2
\mbox{ . }
\label{condition}
\end{equation}
%
%
But looking to the definition of the generalized Fibonacci sequence 
$(F_l)_{l\in {\bf Z}}$ in the
$D$-dimensional case~:
%
%
\begin{equation}
F_{n+1}=\sum_{j=0}^{D-1}F_{n-j} \:\:\: \hbox{with} \:\:\:
F_{j}=0 \:\:\:\mbox{for} \:\:\:j\in [2-D,-1] \:\:\: \mbox{and} \:\:\: 
F_{0}=F_{1}=1
\mbox{ , }
\end{equation}
%
%
one can easily prove, by recursion, that the condition 
(\ref{condition}) is never fulfilled.
This is in contrast with the two-dimensional tilings  originally 
introduced by Rauzy and discussed
in Ref.\cite{Arnoux} that displayed $6$-fold coordinated sites.

%
%
%
%
\section{ Structure Factor}
%
%
%
%

The simple form taken by the codimension 1 tiling sites coordinates
is very helpful to compute the structure factor (which amounts here to the
Fourier transform of the site distribution).
Indeed, the existence of a generating vector allows to write the 
structure factor as a geometric
serie\cite{Mosseri_conumbering2}.
Denoting by $n$ the number of sites in a unit cell and  assuming that 
the sites (atoms) have the
same form factor, the structure factor $S$ reads~:
%
%
\begin{equation}
S({\bf q})={\frac{1}{n}}\sum_{j=0}^{n-1}e^{i\,{\bf q}.{\bf r}^{j}}
\mbox{ , }
\end{equation}
%
%
where ${\bf q}$ is a reciprocal space vector of the approximant 
structure\footnote{
Since approximant structures are periodic, only those vectors contribute to
the structure factor.} and where ${\bf r}^{j}$ represents the vector associated
to the site $j$ (after projection). In the  codimension one case, we 
have seen that
the coordinates were simply expressed in terms of the generator ${\bf g}$~:
%
%
\begin{equation}
{\bf r}^{j}=j\,{\bf g},\;\;j\in [0,n-1]\mbox{ . }
\label{conum_coord}
\end{equation}
%
%
This expression slighly differs from (\ref{babar}) since here, ${\bf 
r}^{j}$ and ${\bf g}$ denotes
the vectors after projection. In addition, the sites are not carried 
back in the unit cell
through the {\it modulo} operation, but since one only considers 
reciprocal space vector, this
does not affect the result. Thus, the structure factor $S$ simply writes~:
%
%
\begin{equation}
S({\bf q})={\frac{1}{n}}\sum_{j=0}^{n-1}e^{i\,j\,{\bf q}.{\bf g}}
\mbox{ . }
\label{serie_geo}
\end{equation}
%
%
The peak intensities (think of a diffraction experiment) are proportional
to~:
%
%
\begin{equation}
|S({\bf q})|={\frac{1}{n}}\left| {\frac{\sin \left( n{\bf q}.{\bf g}
/2\right) }{\sin \left( {\bf q}.{\bf g}/2\right) }}\right| \mbox{ . }
\label{intensite}
\end{equation}
%
%

\begin{figure}[h]
\centerline{\epsfxsize=90mm
\epsffile{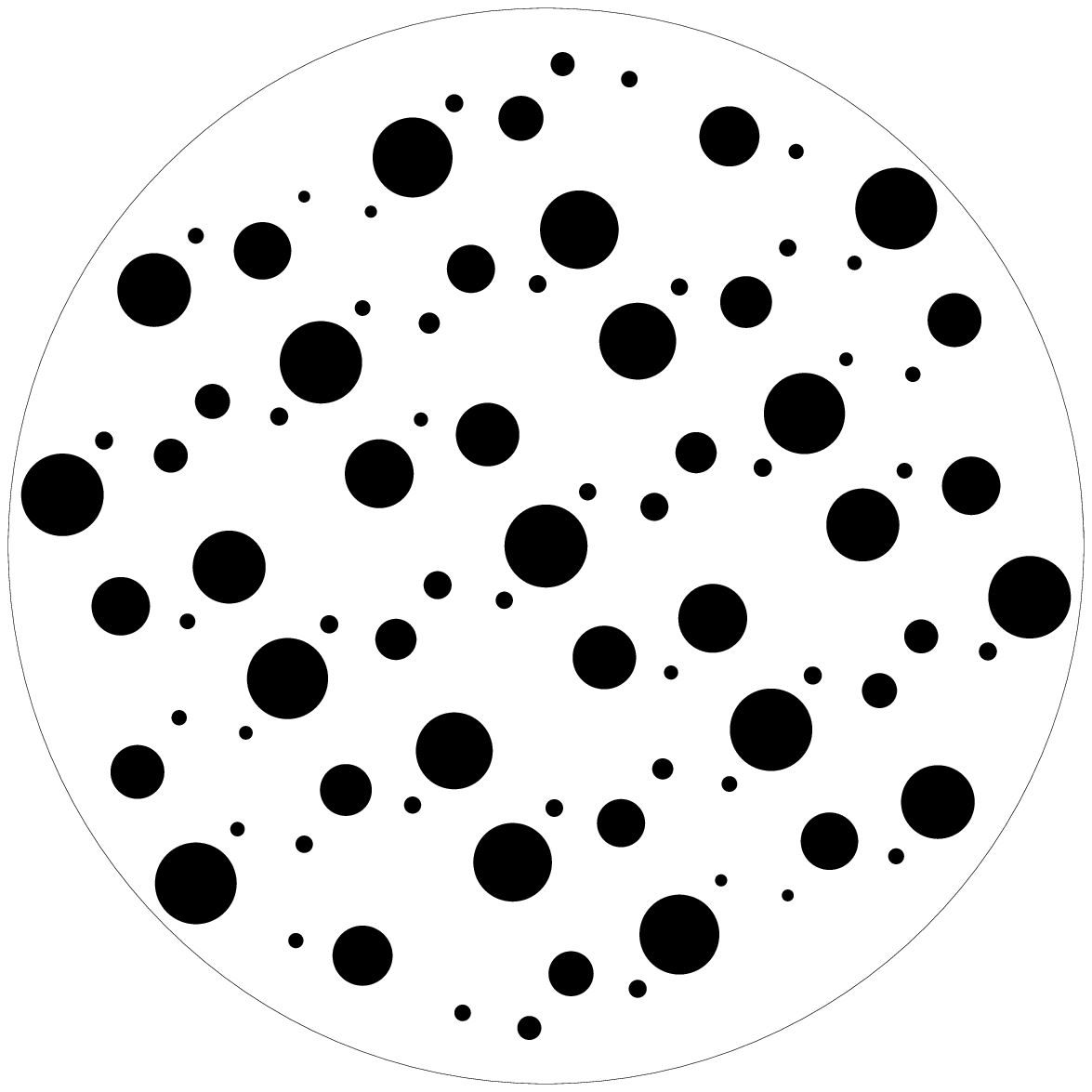}}
\vspace{5mm}
\caption{Structure factor for the $10^{th}$ order approximant (504 
sites per unit
cell). Only intensities greater than an arbitrary threshold are 
shown. The radius
of the circles is proportional to $|S|$.}
\label{TF_Rauzy}
\end{figure}

In the quasiperiodic limit, the reciprocal space unit cell shrinks toward zero,
and the Fourier spectrum becomes dense. However, going from an
approximant to the next one mostly amounts to add new peaks of smaller and
smaller intensities while letting almost unchanged the previous peaks. This
is why any threshold function filtering out the peaks below a certain
(arbitrary) value gives rise to a point-like diffraction pattern (see
figure (\ref{TF_Rauzy}) for an approximant of a two-dimensional 
generalized Rauzy tiling).
%
%
%
%
Note that the Fourier spectrum does not display a particular point-like
rotational symmetry (except the trivial parity ${\bf 
q}\leftrightarrow -{\bf q}$),
in constrast with the more commonly studied octagonal and Penrose
quasiperiodic tilings which have higher codimension (2 and 3 respectively).

%
%
%
%
\section{\protect\bigskip Conclusion}
%
%
%
%
%
The generalized Rauzy tilings form a canonical set of
codimension one quasiperiodic structures. Their construction is based on the
generalized Fibonacci sequence, whose properties allow for a rather 
simple method
to get the sites coordinates as well as their coordination number. We have
shown how to get closed formulas for these quantities valid for
any approximant structures, and which therefore can be carried up to the
quasiperiodic limit. The unidimensional nature of the
perpendicular space allows one to write down the connectivity matrix 
in terms of T{\oe }plitz band-like
matrix, the position of the non vanishing band being directly 
determined from the generalized Fibonacci numbers.

These properties should rank this family of tilings among the most 
interesting to be studied. As  shown
recently\cite{Vidal_ICQ7,Triozon_Rauzy}, they are, to many respect, 
much simpler than the celebrated
Penrose-like tilings, while displaying the same kind of  physical 
properties.\\\\

We would like to thank P. Arnoux, N. Destainville and L. Vuillon for 
very fruitful
discussions about Rauzy original works.\\\\

\appendix
\section{\protect\bigskip One-dimensional generalized Fibonacci 
chains of high codimension}

Up to now, we have described $d$-dimensional codimension one 
structures with a perpendicular
space built from the generalized Fibonacci series. But nothing prevents
to build, along the same line, one-dimensional structures made up of $(d-1)$
different edges, by switching the respecting role of the perpendicular and
parallel space.
Let us briefly describe the obtained sequences and start by
discussing the $3\rightarrow 1$ \ case.

We write down the 3-letter substitution which is given by the matrix $M$,
\[
A\rightarrow AB,\;B\rightarrow AC,\;C\rightarrow A,
\]
which generates (starting from $A$ as an example), the following set of
sequences $S_{l}$:
\[
S_{1}=A,\;S_{2}=AB,\;S_{3}=ABAC,\;S_{4}=ABACABA,\;S_{5}=ABACABAABACAB,\text{%
...}
\]

Note that the length of the $l^{th}$ sequence is given by the
generalized Fibonacci numbers F$_{l}$. The sequences follow a concatenation
rule $S_{l+1}=S_{l}S_{l-1}S_{l-2}$ and the number of occurences of $A$
(resp. $B$, $C$) in the sequence $S_{l}$ is $F_{l-1}$ (resp. 
$F_{l-2}$, $F_{l-3}$). Note that a related sequence
has actually been considered in a previous work\cite{Ali-Gumbs} but 
with the reverse
concatenation rule~: $S_{l+1}=S_{l-2}S_{l-1}S_{l}$.

The generalization for higher codimension is straightforward.



\begin{thebibliography}{10}
\bibitem{Pisotdef} A number, is said to be a Pisot number if it is a root
of a polynomial  with integer coefficients
such that its modulus is bigger than one and all other roots moduli 
are smaller than
one.
\end{thebibliography}
\end{document}